\documentclass[11pt]{article}
\usepackage{amsmath,epsf,amssymb,latexsym,enumerate,cite,shadow,array,color}
\usepackage{slashed}
\usepackage{graphicx,wasysym}

\DeclareFontFamily{U}{rsf}{} \DeclareFontShape{U}{rsf}{m}{n}{
  <5> <6> rsfs5 <7> <8> <9> rsfs7 <10-> rsfs10}{}
\DeclareMathAlphabet\Scr{U}{rsf}{m}{n} \makeatletter
\@addtoreset{equation}{section} \makeatother

\def\be{\begin{equation}}
\def\ee{\end{equation}}
\def\ba{\begin{array}}
\def\ea{\end{array}}

\newcommand{\bea}{\begin{eqnarray}}
\newcommand{\eea}{\end{eqnarray}}

\def\K{K{\"a}hler}
\usepackage{ifpdf}
\ifx\pdfoutput\undefined
   \pdffalse
\else
   \pdfoutput=1
   \pdftrue
  \usepackage[pdftex]{hyperref}
\def\u0{{\underline 0}}

\def\url{{\underline {r+\ell}}}

\newcommand{\rf}[1]{(\ref{#1})}
\newcommand{\vp}{\phi}

\begin{document}

\begin{titlepage}

\hskip 1cm

\vskip 3cm

\begin{center}
{\LARGE \textbf{Inflation, de Sitter Landscape and Super-Higgs effect}}

\

\

  {\bf Renata Kallosh},$^{1}$\,    {\bf  Andrei Linde}$^{1}$   {\bf and  Marco Scalisi}$^{1,2}$
\vskip 0.5cm
{\small\sl\noindent $^{1}$SITP and Department of Physics, Stanford University, Stanford, CA
94305 USA\\
$^{2}$Van Swinderen Institute for Particle Physics and Gravity, University of Groningen,\\
Nijenborgh 4, 9747 AG Groningen, The Netherlands\\
}
\end{center}
\vskip 0.5 cm

\

\begin{abstract}
We continue developing cosmological models involving nilpotent chiral superfields, which provide a simple unified description of inflation and the current acceleration of the universe in the supergravity context. We describe here a general class of models with a positive cosmological constant at the minimum of the potential, such that supersymmetry is spontaneously broken  in the direction of the  nilpotent superfield $S$.  In the unitary gauge, these models have a simple  action where all highly non-linear fermionic terms of the classical Volkov-Akulov action disappear. We present masses for  bosons and fermions in these theories. By a proper choice of parameters in this class of models, one can fit any possible set of the inflationary parameters $n_{s}$ and $r$, a broad range of values of the vacuum energy $V_{0}$, which plays the role of the dark energy, and achieve a controllable level of supersymmetry breaking. This can be done without introducing light moduli, such as Polonyi fields, which often lead to cosmological problems in phenomenological supergravity.
 
 \end{abstract}

\vspace{24pt}
\end{titlepage}


\parskip 7pt

\section{Introduction}

Theoretical physics has to deal with two cosmological surprises: accelerated expansion of the early universe (inflation) and accelerated expansion of the universe now, with $V_0\sim 10^{-120}$ (dark energy). The only presently available explanation of the smallness of $V_0$  is given by the anthropic argument \cite{Linde:1984ir,Weinberg} in the string theory landscape with enormously large set of dS and AdS vacua with different values of $V_{0}$ \cite{Linde:1986fd}.  Supersymmetry provides a certain quality control in the inflationary model construction, and helps to stabilize string theory vacua. Therefore we would like to construct inflationary models which agree with observational data and have an exit into a de Sitter space with many possible values of the positive cosmological constant. Addition of matter and quantum corrections should not change the fact that we have many dS vacua and may again apply the anthropic argument.

The general class of models of inflation in supergravity \cite{Kallosh:2010xz} is constructed using two chiral superfields, with  scalar components
\be
\Phi(x, \theta=0)= {\vp(x) +ia(x) \over \sqrt 2}\, , \qquad S(x, \theta=0)=s(x) \ .
\label{super}\ee
These models are
defined by the following superpotential and \K\, potential
  \be
W= S f(\Phi) \, ,  \qquad K= K ((\Phi-\bar \Phi)^2, S\bar S)\, ,
 \ee
where $f(\Phi)$ was a real holomorphic function, such that $f(\Phi) = \bar f(\bar\Phi)$. The role of the inflaton field in this class of models is played by the field $\phi$. The inflaton potential is $V=|f(\phi/\sqrt 2)|^2$, which allows to  describe  rather general  inflationary models \cite{Linde:1983gd,Linde:2005ht}.  To achieve this in supergravity it was necessary to ensure that the fields $a(x)$ and   $S(x)=s(x)$ vanished during inflation. Usually this happened automatically for the field $a$, whereas the field $S$ often required stabilization, which could be achieved e.g. by the terms $(S\bar S)^{2}$ in the \K\ potential. The minimum of the potential was in Minkowski or de Sitter space with 
\be
V_{0}= |f(0)|^2 \geq 0 \, , \qquad  m^2_{3/2}=|S f(\Phi)|^2 = 0 \ .
 \ee 
The generality of the choice of the function $f(\Phi)$ allows to fit any possible set of the inflationary parameters $n_{s}$ and $r$. As a next step, one should try to describe dark energy/cosmological constant, and also supersymmetry breaking.

In the class of the models considered in  \cite{Kallosh:2010xz} gravitino mass vanishes because it is proportional to $W$, which vanishes throughout the evolution with $S=0$.  In these models the minimum of the potential is at $a(x)=0$ and one can also define the minimum to be at $\Phi(x)=0$, which can be achieved by the redefinition of the field $\phi(x)$. 
 
The supersymmetry breaking terms at the minimum ($\Phi(x)= S(x)=0$) are
 \be
F_S= D_S W= f(0), \qquad F_\Phi =D_{\Phi}W=0\, .
\label{susy}\ee 
In these models  $f(0)=0$ was often considered,  so that  supersymmetric Minkowski vacua were included  in \cite{Kallosh:2010xz}. In particular, in the first and the simplest version of this class of models, the superpotential was chosen to be $W = mS\Phi$ \cite{Kawasaki:2000yn}. Moreover,  any function $f(\phi)$ continuously changing from positive to negative values would automatically produce a Minkowski minimum with unbroken supersymmetry \cite{Kallosh:2010xz}. This was simultaneously an advantage and a limitation: a supersymmetric Minkowski vacuum with unbroken supersymmetry could serve as a good starting point for a small uplifting with $V_0\sim 10^{-120}$ and a small supersymmetry breaking with the gravitino mass in the range of $10^{{-15}} - 10^{{-13}}$ in Planck units, as often required by SUSY phenomenology. One could expect that it is very easy to modify this class of theories to make them fully compatible with phenomenology by introducing some small parameter or making some small changes in the superpotential and \K\ potential. However, it is very difficult to do it because of the no-go theorem described in \cite{Kallosh:2014oja}. One can achieve a tiny uplifting of a supersymmetric Minkowski vacuum to $V_0\sim 10^{-120}$, but it requires a significant change of the model parameters, which typically implies a very strong supersymmetry breaking in the uplifted vacuum. For a recent discussion of a closely related issue see \cite{Linde:2014ela}. 

It is possible to overcome this problem by adding new light scalar fields, such as Polonyi fields, and then solving the cosmological moduli problem related to them, which bothered phenomenologists for more than 30 years \cite{Polonyi}. One can do it, but it requires a special effort with stabilizing the new moduli and making them very massive,  see e.g. \cite{Dudas:2012wi}, and even in this case one should investigate a simultaneous evolution of several scalar fields during inflation.

 Here we will develop a new approach to these issues, building upon  \cite{Kallosh:2010xz} and the recent series of papers on inflation with nilpotent chiral multiplets \cite{Antoniadis:2014oya,Ferrara:2014kva,Kallosh:2014via,Kallosh:2014wsa,Dall'Agata:2014oka}. In order to do it, we will partially change the setting in \cite{Kallosh:2010xz}:

 1. We keep the same \K\, potentials as described in \cite{Kallosh:2010xz},
 \be
 K= K ((\Phi-\bar \Phi)^2, S\bar S). 
 \label{Kmin}\ee
 We preserve the condition  \rf{susy} that supersymmetry is not broken in the inflaton direction. However, we now require that   it is spontaneously broken  in the $S$ direction, with a supersymmetry breaking parameter $M$. Namely, we require that at the minimum of the potential
 \be
F_S= D_S W= f(0)= M\neq 0, \qquad F_\Phi =D_{\Phi}W=0\, .
\label{susy1}\ee 
Thus we exclude the case in \rf{susy} that leads to a supersymmetric Minkowski vacuum,  $W=DW=0$.

 2. We change the status of the $S$ superfield, which will now represent a  Volkov-Akulov theory \cite{Volkov:1973ix} with highly non-linear fermion couplings and spontaneously broken supersymmetry. This is achieved  by promoting the chiral $S$ superfield  to a nilpotent one, defined by $S^2(x, \theta=0)=0$ 
   \cite{rocek,Komargodski:2009rz}.

 3. We add an $S$-independent  function to the superpotential, so that
\be
 W(\Phi, S) = g (\Phi) + Sf(\Phi)\, , \qquad S^2(x, \theta)=0 \ .
\label{newW}\ee
{\it Due to the nilpotency of $S$ and holomorphicity of the superpotential, $W(\Phi, S)$ in eq. \rf{newW}  is the most general form of the superpotential depending on $\Phi$ and $S$}. This is analogous to the fact that an arbitrary function of a single Grassmann variable $\theta$ can be expanded into a Taylor series which terminates after 2 terms, $F(\theta)= a+b \theta$, since $\theta^2=\theta^3=...=\theta^n...=0$. In our case we have $S^2=S^3=...=S^n...=0$.

These three modifications will allow us to preserve the generality of inflationary models but provide a significant flexibility at the minimum of the potential. In our new models $s=0$  during and after inflation, and  we find  the following values of the cosmological constant and gravitino mass in the vacuum:
\be
V_{0} = |f(0)|^2- 3 |g(0)|^2= M^2 - 3  m_{3/2}^2\geq 0\, , \qquad 
  m_{3/2}^2=|W_{\rm min}|^2 = g^2(0)\neq 0  \ .
  \ee
  Here $W_{\rm min}$ is the value of the superpotential at the minimum of the potential, which also defines the mass of gravitino.
 Thus, we need a non-vanishing $g(0)$ to have a massive gravitino, but it should not be too big so that   $  3 |g(0)| ^2 \leq |f(0)|^2$, to avoid a collapsing universe models with a negative cosmological constant  at the end of inflation. In our new models even in case of a Minkowski vacuum, in the limit when $V_{0}\rightarrow 0$, supersymmetry remains broken spontaneously due to \rf{susy1}, with a parameter $M$. This feature is reminiscent of the situation we have found in dS vacua constructed recently in string theory inspired supergravity models in \cite{Kallosh:2014oja}.

We will focus on a class of models which admits  
 a choice of the local supersymmetry gauge  which leads to a simple  version of the generic super-Higgs mechanism \cite{cfgvp,SUGRA}: in general, gravitino `eats' a combination of all other spin 1/2  particles, including  inflatino  and Volkov-Akulov  goldstino and   acquires a mass. The super-Higgs effect in the cosmological setting was studied in \cite{Kallosh:2000ve} where, in particular,  it was shown how the mass matrices are affected by the $\Lambda=3H^2$ in de Sitter space.

The goal of this paper is to develop a theory of the super-Higgs effect in the models with nilpotent fields belonging to the general class of theories described above.  In our models, by design,  {\it gravitino `eats' only the fermion of the nilpotent multiplet,    therefore all complicated Volkov-Akulov type non-linear terms depending on it vanish in the unitary gauge. At the minimum of the potential,  the super-Higgs effect leads to a supergravity action which is simple even in the sector including fermions.  Massive gravitino and inflatino remain in the spectrum and have simple actions.}  On the basis of our results, we will develop a series of models where one can simultaneously describe inflation, dark energy, and supersymmetry breaking without introducing additional scalar fields beyond the scalar part of the inflaton superfield.

The paper is organized as follows. In Section 2 we will briefly review earlier work on related issues. In Section 3 we will describe a general class of models with a nilpotent field $S$ defined by \rf{Kmin}, \rf{susy1} and \rf{newW}. In Section 4 we will develop the theory of the super-Higgs effect in supergravity involving nilpotent chiral superfields, for the class of models described in Section 3. In Section 5 we will examine various mechanisms of uplifting of non-supersymmetric Minkowski vacua to dS vacua in theses theories. In Section 6 we will illustrate our general results by presenting several simple models where one can describe inflation and dark energy, and achieve a controllable level of supersymmetry breaking without introducing additional superfields.

\section{Review of earlier work}

A significant progress in generic inflationary models with the exit to dS minima  was achieved in  \cite{Ferrara:2014kva,Kallosh:2014via}  in the constructions with spontaneously broken Volkov-Akulov (VA) supersymmetry \cite{Volkov:1973ix}. It was also explained in  \cite{Ferrara:2014kva,Kallosh:2014wsa} that VA models have a natural origin in the supersymmetric D-brane physics.
The first Volkov-Akulov Starobinsky supergravity inflation with a nilpotent superfield  with the Minkowski vacuum  was proposed in 
\cite{Antoniadis:2014oya}. 

In \cite{Kallosh:2014via}  some of our models have the property that  at the minimum $F_\Phi = D_\Phi W=0$, supersymmetry is unbroken in the inflaton direction, whereas  $F_S= D_S W\neq 0$, supersymmetry is spontaneously broken in the nilpotent field direction.  In  \cite{Dall'Agata:2014oka} more general models of this type were proposed and studied.
The advantage of such models, as explained in Sec. 2.1 in \cite{Kallosh:2014via}, is that  in this case there is a choice of the local supersymmetry unitary gauge where  all complicated Volkov-Akulov type non-linear terms  vanish. We will return to this issue in Sect. 4 of this paper.

In \cite{Ferrara:2014kva} an underlying superconformal version of constrained chiral superfields, with various constraints, including the nilpotency constraint, was presented.  It was shown, in particular, how to derive the  actions for generic  supergravity models with constrained superfields. The unconstrained off-shell chiral superfield has the form
\be
S(x, \theta)=s(x)+ \sqrt{2}\, \theta\, \chi^s(x) +\  \theta^2 F^S(x) \ ,
\ee
where $s(x)$ is a scalar field,  $\chi^s(x)$ is a fermion partner and $F^S(x)$ is an auxiliary field. It was shown in \cite{Komargodski:2009rz} that the nilpotent superfield\footnote{The relation between the nilpotent superfields and VA models has been recognized long time ago \cite{rocek}.  However, the relevant chiral superfields  in addition to the nilpotency condition $S^2(x, \theta)=0$ were also required to satisfy  an equation of motion. Therefore in cosmological applications we are using the off-shell nilpotent superfields \cite{Komargodski:2009rz}.}
 $S^2(x, \theta)=0$ depends only on the fermion $\chi^s$, VA goldstino,   and an auxiliary field $F^S$, it does not have a fundamental scalar field,  
\be
 S(x, \theta)|_{S^2(x, \theta)=0}  =  \ \frac{\chi^s \chi^s}{2 \, F^S} \ + \ \sqrt{2}\, \theta\, \chi^s \ +\  \theta^2 F^S \ , \label{va1}
\ee
since $s(x)$ is replaced by $ \frac{\chi^s \chi^s}{2 \, F^S}$.
For the nilpotent off-shell superfields  the rules for the bosonic action required for cosmology turned out to be very simple. Namely, one has to calculate potentials as functions of all superfields as usual, and then declare that  the scalar part of the nilpotent superfield $s(x)$ vanishes, since it is replaced by a bilinear combination of the fermions. No need to stabilize and study the evolution of the complex   field $s(x)$.  

However, the inflaton chiral multiplet $\Phi$ contains a complex scalar as shown in eq.  \eqref{super}.
Therefore, in addition to the light inflaton field $\phi(x)$ there is a partner field $a(x)$. An investigation of the models is simplified if this field vanishes during inflation and after it.  This is usually the case, but nevertheless has to be studied for each particular model. Only in case the inflaton $\phi$ belongs to a  vector multiplet, as in  \cite{Ferrara:2013rsa}, the scalar $a$ is absent in the unitary gauge due to a Higgs effect: it is `eaten' by a vector which acquires a  mass.

An important feature of de Sitter vacua in models in \cite{Kallosh:2014via} is that at the minimum
\be
V_0 = F_S \bar F_{\bar S}  - 3 |W_{\rm min}|^2 \ .
\label{Lambda}\ee
Here $F_S= D_S W$ and $\bar F_{\bar S}=\bar D_{\bar S} \bar  W$ and we have taken into account that 
 $K^{\bar S S} =1$ and $K=1$ at the minimum.
Nearly precise balance between the positive goldstino contribution $|F_S|^2$ and the negative gravitino contribution $3 |W_{\rm min}|^2$, along the lines of string landscape scenario, may result in $V_{0}\sim 10^{-120}$. It is important also that in these models there are no light moduli, thus no cosmological moduli problem!

The improved  models of inflation in supergravity with  de Sitter exit discussed in the Introduction will be described below.  Our general models
have a nice feature stressed  also  by Dall'Agata and Zwirner   \cite{Dall'Agata:2014oka}, that the models at the minimum of the potential break supersymmetry spontaneously only in the direction of the nilpotent goldstino multiplet.\footnote{The superpotential  in \cite{Dall'Agata:2014oka} is given by $W = f(\Phi) \, (1 + \sqrt 3 \, S)$. A choice of $\sqrt 3$ is, of course, a well-known  fine-tuning designed to produce $V_{0}=0$. A tiny deviation from $\sqrt 3$ will immediately take the system out of the Minkowski minimum. } Our class of superpotentials depends on two different functions which allow to construct  models with many possible values of the cosmological constant. In fact, the requirement of a strictly Minkowski vacuum at the minimum of the potential, in the classical approximation, does not guarantee that this feature will be preserved when taking into account quantum corrections. Therefore we are looking for  models of inflation with the exit into de Sitter space with all possible values of the cosmological constant,  in the spirit of the  string landscape scenario.

 The models of supergravity interacting with chiral multiplets and vector multiplets are very well known in all details, see for example \cite{SUGRA}. 
One starts with a superconformal action and eliminates auxiliary fields. The superconformal action has a quadratic and linear dependence on all auxiliary fields, therefore by solving equations of motion it is   easy to eliminate them. As the result, the complete bosonic and fermionic supergravity action depends only on physical fields.
Meanwhile, the models of supergravity interacting with Volkov-Akulov fermionic multiplets as well as chiral and vectors multiplets have not yet been constructed, in full generality, in the form in which they depend only on physical fields. It is now known, in principle, how one can derive such models, starting with their superconformal version presented in
\cite{Ferrara:2014kva}. For example, in case of one physical chiral multiplet and one nilpotent multiplet the manifestly supersymmetric action is shown in  eq.  (2.7) in \cite{Ferrara:2014kva}. Since the auxiliary field $F^S$ of the nilpotent multiplet appears in negative powers in fermion-dependent terms such as  ${1\over |F^S|^2} (\chi^s)^2 \partial^2   (\bar \chi^s)^2$, the elimination of $F^S$ will lead to a complicated fermionic part of the classical supergravity action. The bosonic part remains simple, but the fermionic one is complicated. Thus understanding the vacua after the exit from inflation and the  transition to particle physics, reheating and other issues, might not be easy.

The fermion in the VA theory, not interacting with other chiral superfields and gravity, is  a goldstino.  However, in supergravity with chiral multiplets  the goldstino fermion $v$ is a certain combination of all fermions from the chiral and vector multiplets \cite{cfgvp,SUGRA}. This combination is defined by the mixing with gravitino.
Gravitino $\psi^\mu$  interacts with goldstino $v$ as follows:  $\bar \psi^\mu \gamma_\mu \,  v$.

In some models in \cite{Kallosh:2014via},  in models in \cite{Dall'Agata:2014oka}, and in a more general class of models to be described in Sect. 3, where supersymmetry is spontaneously broken in the nilpotent field direction,  $v$ is proportional to a fermion from the nilpotent multiplet $\chi^s$, whereas the inflatino $\chi^\phi$ drops from the generic expression for $v$ due to a condition that $D_\Phi W=0$.
As we will show in Sect. 4, this allows to make a choice of the local supersymmetry gauge $\chi^s=0$, which leads to a simplest possible version of the generic super-Higgs mechanism \cite{cfgvp,SUGRA}.  In such models gravitino `eats'  $\chi^s$,  and therefore all complicated Volkov-Akulov type non-linear terms depending on $\chi^s$ vanish in the unitary gauge. 

\section{General models with VA goldstino `eaten' by gravitino}\label{main}

We start with the simplest \K\, potential in the class considered in \cite{Kallosh:2010xz}, for simplicity, and then our models are defined by
\begin{equation}
K = - \frac12 (\Phi - \bar \Phi)^2 + S\bar S\, ,  \qquad W(\Phi, S) = g (\Phi) + Sf(\Phi)\, , \qquad S^2(x, \theta)=0 \ .
\label{mod}\ee
Our models will describe  rather general   inflationary models \cite{Linde:1983gd,Linde:2005ht} with  the cosmological constant as given by \rf{Lambda}. Our models are designed to
have the following  features:
\be
F_S= D_S W_{\min} =M \, , \qquad F_\Phi= D_\Phi W_{\min}=0  \, , \qquad V_{0} =\Lambda\geq 0  \, ,   \qquad W_{\min} = m_{3/2} \neq 0  \ .
\label{more}\ee
For this purpose we make a choice of both $f$ and $g$ to be real holomorphic functions
\be
\label{real}
{f(\Phi)} = \overline f( \overline{\Phi}) \, , \qquad {g(\Phi)} = \overline g( \overline{\Phi}) \ .
\ee
In terms of our functions in the superpotential eq. \rf{more} means that
\be
 f(0) =M\neq 0\ , \qquad g(0)= m_{3/2}\neq 0 \ .
\label{M+3/2}\ee
We will also impose  a condition
\be
f'(0)=0\, ,  \qquad g'(0)=0 \ ,
\label{V'}\ee
since we want the minimum of the potential to be at $\Phi=0$, for simplicity.

The potential as a function of $(\Phi,\bar{\Phi})$ is
\begin{equation}
\begin{array}{rcl}
V  = 
e^{-\frac{\left(\Phi-\overline{\Phi}\right)^2}{2}} \, \left[ \left| f (\Phi)\right|^2  + \left| g'(\Phi)-g(\Phi) \left(\Phi-\bar{\Phi}\right) \right|^2 -3\left| g(\Phi) \right|^2\right]\,,
\end{array}
\label{ourV}
\end{equation}
where primes stand for derivatives with respect to $\Phi$.
The potential for these models is symmetric under $\Phi\rightarrow\bar{\Phi}$. Thus, in terms of the real components defined by \eqref{super}, the field configurations with $a=0$ and arbitrary $\phi$ are always extrema with respect to $a$. One can identify $\phi$ as the inflaton field, provided that the field $a$ is heavy and is stabilized at $a = 0$. 
During inflation with 
$a=0$ 
\be\label{infpot}
V(\phi)= f ^2\Big ({\phi\over \sqrt 2}\Big )+ \Big( g'\Big ({\phi\over \sqrt 2}\Big ) \Big)^2 - 3 g^2\Big ({\phi\over \sqrt 2}\Big )  \, .
\end{equation}
The minimum of the potential at $\Phi=0$ is defined by an expression 
\be
V'(0) =f(0) f'(0) + g'(0) \Big (g''(0) - 3 g(0) \Big) =0\ ,
\ee
$V'(0)$  vanishes if eq. \rf{V'} is satisfied.

All these models will have a nice and simple fermionic action in the unitary gauge where the non-linearly interacting VA goldstino are `eaten' by gravitino, disappear from the action and make gravitino massive, since we have imposed eq. \rf{M+3/2}.

The potential at the minimum is
\be
V_{0} =\Lambda=f^2(0)- 3 g^2(0) = M^2- 3 \, m_{3/2}^2 \ .
\label{Vmin}\ee
The bosonic masses at the minimum are
\be
m_\phi^2 =   M f''(0)- 3 m_{3/2} g'' (0) + (g''(0) )^2 \ ,
\ee
\be
m_a^2= 2 ( M^2- m^2_{3/2}) 
- M f''(0) - m_{3/2} g''(0)   + (g''(0) )^2  \ .
\ee
The positivity of these masses imposes additional constraints on our models.

We should note that the same potential \rf{infpot} appears in a broad class of theories with \K\ potentials of the functional form $K((\Phi-\bar\Phi)^{2},S\bar S)$, such as, for example, 
 $-3\log \bigl[ 1 +(\Phi -\bar\Phi)^2/6- S \bar S/3\bigr]$. The main difference will be in the expression for the mass of the field $a$ along the inflationary trajectory $a = 0$  \cite{Kallosh:2010xz}. Thus most of the results to be discussed in this paper will remain valid for these more general theories. One may also consider the \K\, potentials $K((\Phi+\bar\Phi)^{2},S\bar S)$. As long as we do not consider interactions of the field $\Phi$ with vectors, these theories are equivalent up to the change of variables.

\section{Fermionic sector after the exit from inflation}
Now we will describe the fermionic sector of the theory.
The generic term mixing  gravitino with  a combination of   fermions from chiral multiplets $\chi^i$,  is proportional to 
\be
 \bar \psi^\mu \gamma_\mu \,  v +h.c= \bar \psi^\mu \gamma_\mu  \sum_i \chi^i e^{K\over 2} D_i W +h.c.
\ee
In case of our two multiplets, in general, both the inflatino $\chi^\phi$ as well as the $S$-multiplet fermion $\chi^s$ form a goldstino $v$, which is mixed with gravitino
\be
 \bar \psi^\mu \gamma_\mu \,  v= \bar \psi^\mu \gamma_\mu  \left( \chi^\phi e^{K\over 2} D_\phi W +  \chi^s e^{K\over 2} D_S W\right)   \ .
\ee
Therefore the local supersymmetry gauge-fixing $v=0$ leads to a condition
\be
v= \chi^\phi e^{K\over 2} D_\phi W +  \chi^s e^{K\over 2} D_S W=0 \ .
\ee
This leads to a mixing of the inflatino $\chi^\phi$ with the $S$-multiplet fermion $\chi^s$. The action has many non-linear in $\chi^s$ terms and therefore the fermionic action in terms of a non-vanishing combination of $\chi^\phi$ and $\chi^s$ is extremely complicated. For example,  a non-gravitational part of the action of the fermion of the nilpotent multiplet  is given by
\be
{\cal L}_{VA}= - M^2 +i \partial_\mu \bar  \chi^s \bar \sigma ^\mu  \chi^s + {1\over 4 M^2}  { (\bar \chi^s)}^2 \partial^2  (\chi^s)^2 - {1\over 16 M^6}  (\chi^s)^2   (\bar \chi^s)^2 \partial^2  (\chi^s)^2 \partial^2  ( \bar\chi^s)^2 \ ,
\label{VA}\ee
as shown in \cite{Komargodski:2009rz}. In  supergravity there will be more non-linear couplings of $\chi^s$ with other fields.

In our class of models where the only direction in which supersymmetry is spontaneously broken is the direction of the nilpotent chiral superfield and $D_\Phi W=0$ the coupling is
\be
\bar \psi^\mu \gamma_\mu  \chi^s e^{K\over 2} D_S W|_{\min} +h.c = \bar \psi^\mu \gamma_\mu \,  \chi^s M +h.c.
\ee
and the goldstino is defined only by one spinor
\be
v= \chi^s M  \ .
\ee
The inflatino $\chi^\phi$,  the spinor from the $\Phi$ multiplet does not couple to $\gamma^\mu \Phi_\mu$ since $D_\Phi W|_{\min}=0$. In this case we can make a choice of the unitary gauge  $v=0$, when fixing local supersymmetry.  Since $M\neq 0$  it means that we can remove the spinor from the nilpotent multiplet
\be
\chi^s=0 \ .
\ee
The corresponding gauge is the one where gravitino becomes massive by `eating' a goldstino. The unitary gauge is a gauge where the massive gravitino has both $\pm 3/2$  as well as $\pm1/2$ helicity states.
In our models the fully non-linear fermion action simplifies significantly since it depends only on inflatino. All complicated non-linear terms of the form 
$
{1\over M^2} (\chi^s)^2 \partial^2   (\bar \chi^s)^2
$ and higher power of fermions as well as mixing of the inflatino  $\chi^\phi$ with $\chi^s$ disappear  in this unitary gauge. 

In particular, the fermion masses of gravitino and inflatino,  at the minimum are simple
\be
m_{3/2}=W_{0}=g(0) \ , \qquad m_{\chi^\phi}= e^{K\over 2}\partial_\Phi D_\Phi W = g''(0)- g(0) = g''(0)- m_{3/2} \ .
\ee
Here we have presented the masses of fermions without taking into account the subtleties of the definition of such masses in the de Sitter background. This was explained in details for spin 1/2 and spin 3/2 in  \cite{Kallosh:2000ve}  in case including  $\Lambda>0$. For example, the chiral fermion mass matrix ${\bf m}^{ij}= D^i D^j e^{K\over 2} W$  is replaced by
$
\hat m \equiv {\bf m} + \sqrt {\Lambda/3}\, \gamma_0 
$.

\section{General models with $\Lambda\approx 0$}
In our practical applications we will be mostly interested in the models which can describe the observable non-zero but extremely small value of the cosmological constant $V_{0} \sim 10^{{-120}}$. Of course, it is hopeless to calculate $V$ with such an enormous accuracy, but for our purposes it is sufficient that the result will be sufficiently small, 
\be
V_{0} =\Lambda=f^2(0)- 3 g^2(0) = M^2- 3 \, m_{3/2}^2\ll M^{2} \ .
\label{0}\ee
Then, following the ideas of the string landscape scenario, we may argue that among many string theory vacua there will be some where the effective parameters of the theory are such that the resulting vacuum energy will take its observational value, which is required for the existence of life as we know it.

As a starting point, one may try to find the models of our type with a Minkowski vacuum with $V_{0} = 0$ and spontaneously broken supersymmetry. This is achieved for $M^2- 3 \, m_{3/2}^2=0$, i.e. for
\be
m_{3/2}=\pm  M/\sqrt 3 \not = 0 \ .
\ee
This will be the first step of our investigation. After that, we will show how one can uplift such models to a locally stable dS state with a small value of the cosmological constant. The possibility to do it is in fact quite nontrivial. In the previously studied class of models with a stable {\it supersymmetric} Minkowski vacuum \cite{Kawasaki:2000yn,Kallosh:2010xz}, with a positively defined mass matrix, one would be unable to uplift this vacuum to a dS vacuum with a small SUSY breaking by an infinitesimal change of the \K\ potential and the superpotential without introducing new superfields, because of the no-go theorem formulated in \cite{Kallosh:2014oja}. Fortunately, if one finds a {\it non-supersymmetric} Minkowski vacuum with $m_{3/2} \not = 0$, its uplifting to dS becomes possible without introduction of extra scalar fields, such as light Polonyi fields, as we are going to demonstrate below.  If the uplifting is sufficiently small, it will not affect any consequences of the inflationary regime in the models with $V_{0} = 0$. 

The bosonic masses at the minimum are
\be
m_\phi^2 =   M f''(0)\mp  \sqrt 3 M g''(0) + (g''(0))^2 \ ,
\ee
\be
m_a^2= {4\over 3} M^2
- M f''(0) \mp  M g''(0)/\sqrt 3  + (g''(0))^2  \ .
\ee
Gravitino and inflatino have the following masses
\be
m_{3/2}=\pm  M/\sqrt 3 \ , \qquad m_{\chi^\phi} = g''(0)\mp  M/\sqrt 3  \ .
\ee

For simplicity of the model building, it may be convenient to represent the general class of models  \rf{mod} in such a way that the main constraint $f'(0)=g'(0)=0$ \rf{V'} is satisfied automatically. One can do it by replacing the functions $g(\Phi)$ and $f(\Phi)$ by  $\tilde g(\Phi) -\Phi\, \tilde g'(0)$ and $\tilde f(\Phi)  -\Phi\, \tilde f'(0)$:
\be
W = \tilde g(\Phi) -\Phi\, \tilde g'(0) + S\left (\tilde f(\Phi)  -\Phi\, \tilde f'(0)\right) \ .
\label{subtr}\ee
Now one can use arbitrary functions $\tilde g(\Phi)$ and $\tilde f(\Phi)$; the required conditions $f'(0)=g'(0)=0$ will be satisfied.

As a next step, one can further modify the choice of the functions $f(\Phi)$ which will automatically bring us to the Minkowski vacuum state with $V_{0} =M^2- 3 \, m_{3/2}^2 = 0$. This can be done by one of the two different choices, which simultaneously keep track of the conditions $f'(0)=g'(0)=0$: $f(\phi) = \tilde f(\Phi) -\tilde f(0) -\Phi\, \tilde f'(0) \pm \sqrt 3 \tilde g(0)$:
\be
W = \tilde g(\Phi) -\Phi\, \tilde g'(0) + S\left (\tilde f(\Phi) -\tilde f(0) -\Phi\, \tilde f'(0) \pm \sqrt 3 \tilde g(0)\right) \ .
\label{subtr2}\ee
One can show that this general superpotential does indeed lead to the Minkowski vacuum state with $V_{0} =0$, for any choice of the  functions $\tilde g(\Phi)$ and $\tilde f(\Phi)$.

There are three slightly different ways to uplift the Minkowski vacuum to a dS vacuum. 

1) One can add a small constant $d$ to $f(\Phi)$:
\be
W = \tilde g(\Phi) -\Phi\, \tilde g'(0) + S\left (\tilde f(\Phi) -\tilde f(0) -\Phi\, \tilde f'(0) \pm \sqrt 3 \tilde g(0)+d\right) \ .
\label{uplift}\ee
After that, the value of the potential in the minimum becomes
\be
V_{0} =d\, (d\pm 2\sqrt 3 g(0)) \ .
\label{ds}
\ee
By a proper choice of $d$, one can uplift the extremum of the potential at $\Phi = 0$ to any positive value of $V_{0}$. 

2) One can add a small constant $d$ to $g(\Phi)$:
\be
W = \tilde g(\Phi) -\Phi\, \tilde g'(0) +d+ S\left (\tilde f(\Phi) -\tilde f(0) -\Phi\, \tilde f'(0) \pm \sqrt 3 \tilde g(0)\right) \ .
\label{uplift}\ee
After that, the value of the potential in the minimum becomes
\be
V_{0} = -3 d (d + 2 g(0)) \ .
\label{ds}
\ee
Notice the change of sign in front of the term $d^{2}$. By a proper choice of $d$, one can downshift the extremum of the potential at $\Phi = 0$ to any negative value of $V_{0}$, or uplift it by some value in the range from $0$ to $3g^{2}(0)$. 

3) One can multiply $f(\Phi)$ by $(1+d)$:
\be
W = \tilde g(\Phi) -\Phi\, \tilde g'(0) + S\left (\tilde f(\Phi) -\tilde f(0) -\Phi\, \tilde f'(0) \pm \sqrt 3 \tilde g(0)\right)(1+d) \ .
\label{uplift}\ee
After that, the value of the potential in the minimum becomes
\be
V_{0} =3d\, (2+d) g^{2}(0) \ .
\label{ds}
\ee
By a proper choice of $d$, one can uplift the extremum of the potential at $\Phi = 0$ to any positive value of $V_{0}$. 

Of course, in any particular model one should also verify that this extremum after the uplifting remains a stable minimum of the potential for the physically interesting choice of the parameters.

In the following section we will show how specific choices for the functions $g(\Phi)$ and $f(\Phi)$ can realize the three uplifting mechanisms described above while assuring working inflation models. We will see how the general class discussed in Sec.~\ref{main} allows for a broad variety of scenarios while including some already discussed in literature.

\section{New models $W= g(\Phi) + Sf(\Phi)$ in examples}
\subsection{ Models with $g(\Phi)= W_0 = const$}

The simplest model of this class
was presented in \cite{Kallosh:2014via} in Sec. 2.1. It was the first example of the cutting off non-linear fermions from the spectrum in the context of the inflation with de Sitter exit models. It was described by 
\be
K=-{(\Phi- \bar\Phi)^2\over 2} + S\bar S  , \qquad 
 W=  SM(1+c \Phi^{2}) +W_{0}\ .
\label{1ii}\ee
Here the constant $W_{0}$ should not be confused with $W_{\min} = g(0) = m_{{3/2}}$. This model corresponds to the following choice of the functions $g(\Phi)$ and $f(\Phi)$ in \rf{mod}:
\be
g(\Phi)= W_{0}\, , \qquad f(\Phi) =  M(1+c \Phi^{2}) \ .
\label{1ii}\ee
 The potential is 
\be
V(\phi,a) ={e^{\,a^{2}}}  \Bigl[M^{2}(1+c\, (\phi^{2}-a^{2})) +W_{0}^{2}\, (2a^{2}-3)+{M^{2}c^{2}\over 4}(\phi^{2}+a^{2})^{2}\Bigr]\ .
\label{pot}\ee
 The calculation of the mass squared of the field $a$ during inflation in this scenario for $c \ll 1$ shows that it is greater than $6H^{2}$, so the state $a =0$ is stable. During inflation one has $a = 0$, and the inflaton potential becomes 
\be\label{quad}
V(\phi) =cM^{2} \phi^{2}\ \Bigl(1 +{c\over 4}\phi^{2}\Bigr) +V_{0} \ .
\ee
For $c\ll 1$, the potential remains nearly exactly quadratic during the last 60 e-foldings of inflation. The value of  the potential in its minimum is given by
$
V_{0} =M^{2} -3W_{0}^{2} 
$.
 The state $a = 0$ remains stable after inflation as well.
 
For comparison with other models to be discussed in this paper, one may represent $M$ as $b g(0)$. Then the expression for the vacuum energy will look as follows:
\be
V_{0}= g^{2}(0)\, (b^{2} -3) \ .
\ee
The vacuum energy disappears for $b = \sqrt 3$. We encounter similar expressions in all models considered in our paper, as well as in   \cite{Kallosh:2014via,Dall'Agata:2014oka}. The observable value of $V_{0}$ is about $10^{-120}$. Our main goal is not to describe the models with $b = \sqrt 3$ and $V_{0}= 0$, but the realistic models with a non-zero value of the cosmological constant. In many supergravity models, the minuscule uplifting by $V_{0} \sim 10^{{-120}}$ represents a significant problem, but in the models with nilpotent fields this problem is easily solved by a minor variation of parameters such as the parameter $b$. 

While the model \rf{1ii} is internally consistent, it has a very large supersymmetry breaking  with $m_{3/2}\gtrsim 2.5\times 10^{-4}$  \cite{Kallosh:2014via}. Such models may gain in popularity if no sign of supersymmetry is discovered at LHC.  To pursue a more often discussed possibility of the low scale supersymmetry breaking with $m_{3/2} \sim 10^{-15} - 10^{-13}$ in Planck units, we will proceed with more general choices of functions $g(\Phi)$ and $f(\Phi)$.

\subsection{Models with $ g(\Phi) ={ f(\Phi) \over b}$ }

As a next step, one may consider models with $f(\Phi)= b\,  g(\Phi)$, where $b$ is some constant. This implies the following choice of the superpotential: 
\begin{equation}
W = g(\Phi) \, (1 + b\, S) \, , 
\label{M}\end{equation}
where 
\be
g^{\, \prime} ( 0 ) = 0 \, , 
\qquad
 b \, g( 0 ) =f(0)= M \ne 0 \, .
\end{equation}
 
 For $b = \sqrt 3$, this model is equivalent to the model proposed by Dall'Agata and Zwirner  \cite{Dall'Agata:2014oka}, but in our formulation, following \cite{Kallosh:2010xz}, the inflaton field is a real part $\phi$ of the complex scalar, which makes this formulation a bit more intuitive.

The inflaton potential for $a = 0$ is given by
\begin{equation}
\label{Vinf}
V_{\rm inf}  =(b^2-3)\, g ^2\Big ({\phi\over \sqrt 2}\Big )+ \Big( g'\Big ({\phi\over \sqrt 2}\Big ) \Big)^2  \, .
\end{equation}
Stability of the field $a$ is determined by the mass of this field along the inflationary trajectory $a =0$:
\be
m_{a}^2= 2(b^2-1)g ^2\Big ({\phi\over \sqrt 2}\Big )  +(3+b^2)\Big( g'\Big ({\phi\over \sqrt 2}\Big ) \Big)^2 -(1+b^2) g \Big ({\phi\over \sqrt 2}\Big )g''\Big ({\phi\over \sqrt 2}\Big )+\Big(g''\Big ({\phi\over \sqrt 2}\Big ) \Big)^2 - g'\Big ({\phi\over \sqrt 2}\Big ) g'''\Big ({\phi\over \sqrt 2}\Big )\,.
\ee

One could expect that during inflation $m^{2}_{a} \sim H^{2} \sim V/3$, as often happens with generic scalar fields in an expanding universe. However, in this class of models the relation is different. In the limiting case  $b^{2} -3\ll 1$, the term $g^2\Big ({\phi\over \sqrt 2}\Big )$ disappears from the expression for the inflaton potential, which becomes
\be
\label{Vinf2}
V_{\rm inf}   \approx \Big( g'\Big ({\phi\over \sqrt 2}\Big ) \Big)^2 \, .
\end{equation}
However, the term $g^2\Big ({\phi\over \sqrt 2}\Big )$ does not disappear from the expression for $m^{2}_{a}$.  During inflation at $\phi \gg 1$, $b^{2} -3\ll 1$, one typically has 
\be
m_{a}^2 \approx 4 g ^2\Big ({\phi\over \sqrt 2}\Big ) \gg H^{2} \approx {1\over 3}\Big( g'\Big ({\phi\over \sqrt 2}\Big ) \Big)^2.
\ee
Thus the stabilization of the field $a$ in this class of theories is even stronger than expected, so the field $a$ very rapidly disappears during inflation. This unusual result typically remains true for other scalar fields which one may incorporate into this model. 

After inflation, when the field reaches the vacuum at $ \Phi=0$, we have
\be
\label{V0}
V_{0}  =  \Big (1-{3\over b^2} \Big )  M^2= M^2- 3 m_{3/2}^2\, .
\end{equation}
as in models in \cite{Kallosh:2014wsa}.
Thus we find that 
\begin{equation}
D_S W_{0} = b g(0)= f(0)= M \ne 0 \, , 
\qquad
D_\Phi W_{0} = {g}^{\, \prime} (0) = 0 \, , 
\qquad
m_{3/2} =g(0) = {M\over b} \, .  
\end{equation}
The scalar masses are
\be
m_\phi^2 = (g''(0))^2 + (b^2-3) g(0)g''(0)\, , \qquad 
m_a^2= (g''(0))^2- (1+b^2) g(0)g''(0) +2(b^2-1) g(0)^2 \ .
\ee

The fermion masses of gravitino and inflatino at the minimum are simple
\be
m_{3/2}=W_{\min}={M\over b} \ , \qquad m_{\chi^\phi}= e^{K\over 2}\partial_\Phi D_\Phi W = g''(0)- g(0) \ .
\ee
In particular,  we will be  interested in the limit  $b^{2}-3 \ll 1$, which may describe the  case when $V_{0} \sim 10^{{-120}}$. In this limit
\begin{equation}
\label{masses}
m^2_\phi= (g''(0))^2 \, , 
\qquad
\qquad
m^2_a =  ( g^{\, \prime \prime} (0) - 2 \, M/b)^2 \, .
\end{equation}
The fermion masses of gravitino and inflatino in the (nearly) Minkowski minimum are 
\be
m_{3/2}=W_{\rm min}={M\over \sqrt 3} \ , \qquad m_{\chi^\phi}=  g''(0)- {M\over \sqrt 3} \ .
\ee
Importantly, the parameters $m{3/2}$ and $M$, describing the amplitude of supersymmetry breaking after inflation, are not directly related to the description of inflation in this scenario. As a result, in this class of models, unlike in the model described in the previous section, one can have large scale of inflation and small scale of supersymmetry breaking. We will illustrate it now by considering some simple examples.

\subsubsection{A simple example  with $ g(\Phi) =\lambda+{m\over 2} \Phi^{2}$ }
The simplest version of the model \rf{M} with $ g(\Phi) ={ f(\Phi) \over b}$ is the model with $f(\Phi) = b(\lambda+{m\over 2} \Phi^{2})$:
\be
W = \left(\lambda+{m\over 2} \Phi^{2}\right) \, (1 + b\, S) \ .
\ee
For $b = \sqrt 3$ this model was described in \cite{Dall'Agata:2014oka}.  Here we will consider general values of $b$, which should help us to describe a broad range of values of the cosmological constant, i.e. simultaneously describe inflation and dark energy using this simple supergravity model. 

The potential at its minimum at $\phi = a =0$ is given by
\be
V_{0} = M^2- 3 m_{3/2}^2 = (b^{2}-3)\lambda^{2} \ .
\ee
During inflation, the tiny deviation of $V_{0}$ from zero is unimportant, so at that stage one can simplify all equations taking the limit $b^{2}-3\ll 1$. In that case, the inflaton potential at $a = 0$ is 
\be
V(\phi) = {m^{2}\over 2}\phi^{2} \ ,
\ee
and the mass of the field $a$ during inflation with $\phi \gg 1$ is
\be
m^{2}_{a} = {m^{2}\over 4} \phi^{4} \ .
\ee

In this model we have three different mass/energy scales. First of all, the inflaton mass $m \sim 10^{{-5}}$ is responsible for the large energy density scale during inflation. Then, by tuning the mass scale $m_{3/2} \approx {M\over \sqrt 3} \ll m$, one can describe a much smaller mass scale corresponding  to supersymmetry breaking. Finally,  due to a partial cancellation between the parameters $b^{2}\lambda$ and $3\lambda$ (i.e. between $M^2$ and  $3 m_{3/2}^2$) one can obtain any desirable value of the cosmological constant, including the desirable value $V_{0} \sim 10^{{-120}}$, which allows to describe dark energy. 

As we already mentioned, during inflation, for $\phi \gg 1$, one has $m^{2}_{a} \gg H^{2}$, so the field $a$ rapidly falls down towards $a = 0$ and stays there. One may even wonder whether $m^{2}_{a}$ under certain conditions may become too large. Indeed, for $\phi \gtrsim m^{{-1/2}}$ this mass becomes greater than the Planck mass, and the whole calculation goes out of control. Fortunately, this does not affect the observational consequences of the model, since for the realistic value $m \sim 6\times 10^{{-6}}$ the boundary at $\phi \gtrsim m^{{-1/2}}$ corresponds to the unobservable region with the number of e-foldings $N \sim 1/m \sim 10^{5}$. 

Interestingly, this boundary coincides with the boundary of eternal inflation regime in this theory, which occurs for $\phi \gtrsim m^{{-1/2}}$    \cite{Linde:2005ht}. Even more interestingly, the same conclusion is valid for other models with the potentials $\phi^{n}$. Thus in this class of inflationary models there may be no slow-roll eternal inflation. However, this conclusion is valid only for the simplest class of models where $m^{2}_{a}$ can approach Planckian values. There are many models where it never happens and a slow-roll eternal inflation may occur. Such models include hilltop inflation, Starobinsky model, universal attractors and many other large field models. The problem of initial conditions for inflation in such models can be solved by the methods outlined in \cite{Linde:2014nna}. We should also note that the regime of eternal inflation may always exist due to tunneling between different metastable dS vacua in the landscape.

\subsection{Models with $g(\Phi) = W_0 + {f(\Phi)\over b}$ }
Here is an example of  different models  of the kind described in \rf{mod}, \rf{more}:
\be\label{C}
W= W_0  +  {f(\Phi)\over b}+ f(\Phi) S \ ,
\ee
which means that we have chosen
\be
g(\Phi)= W_0  +  {f(\Phi)\over b} \ .
\ee

The inflaton potential at $a =0$ is given by 
\be
V(\phi)= f ^2\Big ({\phi\over \sqrt 2}\Big )+  b^{-2}\Big( f'\Big ({\phi\over \sqrt 2}\Big ) \Big)^2 - 3 \left( W_0  +  b^{{-1}}\, f\Big ({\phi\over \sqrt 2}\Big )\right)^2   \, .
\end{equation}
The potential at the minimum with $m_{3/2}= W_0  +  f(0)/b$ is
\be
V_{0} =\Lambda=f^2(0)- 3 (W_0  +  f(0)/b)^2 = M^2- 3 \, m_{3/2}^2 \ .
\label{Vmin}\ee
It vanishes if $M^2- 3 \, m_{3/2}^2
=0$, which happens if
\be\label{zero}
W_0 = - f(0)\left({1\over b} \pm {1\over \sqrt 3}\right) \ .
\ee
Thus for  $b^2= 3$  one should have  $W_0=0$ to obtain a Minkowski vacuum, as in the previous subsection. However, one can have $V_{0} = 0$ for other values of $b$ as well, if $W_0$ is given by  \rf{zero}.  

More importantly, by changing the parameters $b$ and $W_0$ one can find change the value of $V_{0}$ in a broad range of its values. Thus, one can find inflationary theories with dS minima in a rather general class of superpotentials, which is good from the point of view of the possibility of the further generalizations of this class of models.

\subsubsection{A simple example  with  $f(\Phi) = b(\lambda+{m\over 2} \Phi^{2})$ }

Now we will consider the simplest  model in \rf{C}  with $f(\Phi) = b(\lambda+{m\over 2} \Phi^{2})$ which corresponds to the superpotential
\be
W = W_0+ \left(\lambda+{m\over 2} \Phi^{2}\right) \, (1 + b\, S) \ .
\ee
The inflaton potential at $a =0$ in this model is given by 
\be
V(\phi)= V_{0} +{\phi^2\over 2} \left(m^2 -3 W_0 m  + (b^2-3) \lambda m\right) + {  \phi^4\over 16} (b^{2}-3)    \, .
\ee
The potential at the minimum with $m_{3/2}= \lambda +W_0$ is
\be
V_{0} =b^2\lambda^2 -3 (W_0+\lambda)^{2} \ .
\label{Vmin11}\ee
$V_{0}$ vanishes if $M^2- 3 \, m_{3/2}^2
=0$, which happens for
\be\label{zero1}
b^2 = 3 (1+W_0 \lambda^{{-1}})^{2} \ .
\ee
By increasing $W_0$ in \rf{Vmin11} one can make the cosmological constant negative, whereas by increasing $b$ one can make it positive.

\

\section{Conclusions}
We  investigated various  features of cosmological models involving nilpotent chiral multiplets in supergravity. In models with one chiral multiplet $\Phi$ and one nilpotent multiplet $S$ the most general superpotential is defined by two functions, $W= g(\Phi) + S f(\Phi)$. We consider models with \K\, potentials of the type $K= K ((\Phi-\bar \Phi)^2, S\bar S)$. A bosonic part of the action for these models is very simple and by now it is well understood. This may be sufficient for the description of physical processes during inflation. Moreover, the theories with nilpotent multiplets offer numerous advantages for investigation of supersymmetry breaking and uplifting of Minkowski vacuum to a dS vacuum.

However, the fermionic part in these models, in general, is quite complicated and requires special investigation, which is necessary to calculate the fermion masses, study their interactions and ensure the internal consistency of the theory.  In this paper we described a broad class of models of this type where $D_{\Phi} W=0$ and $D_S W=M \not = 0$ in the minimum of the scalar potential. In these models gravitino is mixed only with the goldstino from the nilpotent multiplet. Because of that, the supergravity action in the unitary gauge  depends on gravitino and inflatino, and it is simple. The complicated non-linear dependence on goldstino, which is present in the classical action, disappears in this gauge due to the super-Higgs effect. In this class of models one can calculate masses of all bosons and fermions, describe inflation and the present stage of acceleration of the universe, and achieve a controllable level of supersymmetry breaking without introducing any additional superfields such as Polonyi fields.

\section*{Acknowledgments}

We are grateful to S. Ferrara, K. Dasgupta, S. Kachru, J. Maldacena, N. Seiberg and  T. Wrase for important discussions. RK and AL are supported by the SITP and by the NSF Grant PHY-1316699. RK is also supported by the Templeton foundation grant `Quantum Gravity Frontiers,' and AL is supported by the Templeton foundation grant `Inflation, the Multiverse, and Holography.' MS acknowledges financial support by the University of Groningen, within the Marco Polo Fund, and by COST Action MP1210. Finally, MS would like to thank the SITP for its warm hospitality.

\

\end{document}